%% file: main.tex
\documentclass{article}[12pt]
\usepackage[utf8]{inputenc}
 \def\trans{^{\transpose}}
 \def\bbeta{\boldsymbol{\beta}}

\newcommand\independent{\protect\mathpalette{\protect\independenT}{\perp}}
\def\independenT#1#2{\mathrel{\rlap{$#1#2$}\mkern2mu{#1#2}}}

\def\sone{^{(1)}}
\def\szero{^{(0)}}
\def\sa{^{(a)}}

\def\bXcheck{\check{\bX}}
\def\bScheck{\check{\bS}}
\def\Acheck{\check{A}}

\def\Deltacheck{\check{\Delta}}
\def\Ebb{\mathbb{E}}

\usepackage{amsthm}
\newtheorem{remark}{Remark}
\newtheorem{proposition}{Proposition}
\usepackage{graphicx,verbatim,array,multicol, subfigure, color, lscape, mathrsfs}
\usepackage{psfrag, amsmath, amsfonts, epsfig, fancybox, setspace,soul, amsthm}
\usepackage{longtable}

\def\sone{^{(1)}} 
\def\szero{^{(0)}}


\input{GrandMacros.tex}

\title{Doubly-robust evaluation of high-dimensional surrogate markers}
\author{Denis Agniel, Layla Parast, Boris Hejblum}

\begin{document}
\maketitle
\begin{abstract}
    When evaluating the effectiveness of a treatment, policy, or intervention, the desired measure of effectiveness may be expensive to collect, not routinely available, or may take a long time to occur. In these cases, it is sometimes possible to identify a surrogate outcome that can more easily/quickly/cheaply capture the effect of interest. Theory and methods for evaluating the strength of surrogate markers have been well studied in the context of a single surrogate marker measured in the course of a randomized clinical study. However, methods are lacking for quantifying the utility of surrogate markers when the dimension of the surrogate grows and/or when study data are observational. We propose an efficient nonparametric method for evaluating high-dimensional surrogate markers in studies where the treatment need not be randomized. Our approach draws on a connection between quantifying the utility of a surrogate marker and the most fundamental tools of causal inference -- namely, methods for estimating the average treatment effect. We show that recently developed methods for incorporating machine learning methods into the estimation of average treatment effects can be used for evaluating surrogate markers. This allows us to derive limiting asymptotic distributions for key quantities, and we demonstrate their good performance in simulation.
\end{abstract}
\section{Introduction}
When evaluating the effectiveness of a treatment, policy, or intervention, the desired measure of effectiveness may be expensive to collect, not routinely available, or may take a long time to occur. In these cases, it is sometimes possible to identify a surrogate outcome that can more easily/quickly/cheaply capture the effect of interest. For example, when evaluating an intervention designed to delay dementia onset, the time required to observe enough dementia diagnoses is often very long and surrogates that have been considered for intervention evaluation include mild cognitive impairment, adiponectin levels, neuroimages, amyloid plaques and neurofibrillary tangles \cite{small2006diagnostic,kantarci2004quantitative,teixeira2013decreased,levey2006mild,mathis2007impact}. Similarly, in studies evaluating treatments to prevent diabetes, surrogate measures for diabetes onset have included changes in body weight, fasting plasma glucose and hemoglobin A1c \cite{caveney2011diabetes,diabetes200910,choi2011hemoglobin,Parast2017}. 

Theory and methods for evaluating the strength of surrogate markers have been well studied in the context of a single surrogate marker measured in the course of a randomized clinical study. In particular, robust nonparametric methods have been proposed in \cite{parast2019assessing, parast2016robust}. However, fully nonparametric methods are not available or not reliable when the numbers of markers is more than one or two. In these cases, an initial model may be used to reduce the dimensionality of the surrogate \cite{agniel2020evaluation,parastevaluating,parast2016robust}, but an inappropriate initial model can produce badly biased estimates of the utility of the surrogate. Additionally, as the dimension of the surrogate nears or exceeds the sample size, initial parametric models may not have adequate properties or might not even be possible. Furthermore, assessments of surrogate markers are not solely limited to studies of randomized treatment \cite{de2004measurement,obirikorang2012total}. In fact, using surrogate outcomes based on complex observational data is finding purchase in all corners of science \cite{wang2020methods}. 

In this work, we propose an efficient nonparametric method for evaluating high-dimensional surrogate markers in studies where the treatment need not be randomized. Our approach draws on a connection between quantifying the utility of a surrogate marker and the most fundamental tools of causal inference -- namely, methods for estimating the average treatment effect. We take advantage of state-of-the-art methods for incorporating flexible machine learning and/or sparse high-dimensional models into estimation of treatment effects via the cross-fitting approach of \cite{Chernozhukov2017}. Cross-fitting allows us to derive asymptotic distributions and compute confidence intervals for key quantities while putting minimal restrictions on what types of estimators are used. Furthermore, our estimator and results may be used regardless of the dimension of the surrogates.

The structure of the rest of the paper is as follows. In Section 2, we lay out notation and the setting in which we are working, and we motivate the importance of evaluating surrogate markers in light of recent advances in estimating surrogates from high-dimensional data. In Section 3, we detail assumptions necessary for identifying and interpreting parameters of interest, as well as our approach to estimating the quality of a surrogate marker. We derive the asymptotic behavior of our estimator in Section 4 and discuss variance estimation and inference. We evaluate the performance of the proposed approach using a simulation study in Section 5
. We give final remarks and draw connections to other methods for evaluating surrogate markers in Section 6.

\section{Evaluation of Surrogate Markers}

\subsection{Notation}
Let $A$ denote a binary treatment, and let the primary outcome of the study be $Y$. Let there be a vector $\bS$ of potential surrogate information, and let $\bX$ be a vector of pre-treatment covariates. The primary quantity of interest is the treatment effect on the outcome 
\[
\Delta = \Ebb\{ Y\sone - Y\szero \}
\]
where $Y\sa$ is the potential or counterfactual outcome that would have been observed if the observed value of treatment were $A = a$, possibly contrary to fact. Similarly let $\bS\sa$ be the potential/counterfactual value the vector of surrogates would take if $A = a$. We desire to approximate $\Delta$ via a difference in some function of the surrogates $\bS$ and covariates $\bX$. Let $\psi_a(\bX, \bS)$ be such a scalar function that is here indexed by treatment $a$ but need not depend on treatment. Let the data observed in the current study be $\Osc = (\bX_i, A_i, \bS_i, Y_i)_{i=1,...,n}$, $n$ iid realizations of $\bO = (\bX, A, \bS, Y)$. And let $\Fsc = (\bXcheck_i, \Acheck_i, \bScheck_i)_{i=1,...,n^*}$ be $n^*$ iid realization of $\bO$ in a future study. The function $\psi_a(\bx, \bs)$ is useful and is of particular interest because for a future study, one could then use $\psi_i = \psi_1(\bXcheck_i, \bScheck_i)I\{\Acheck_i = 1\} + \psi_0(\bXcheck_i, \bScheck_i)I\{\Acheck_i = 0\}$ as the outcome instead of $Y$. The true treatment effect $\Delta$ could be approximated by estimating the effect of $\Acheck_i$ on $\psi_i$ conditional on $\bXcheck_i$ via regression, weighting, matching, or any other reasonable approach.

\subsection{Importance of quantifying surrogate strength}

Understanding the strength of a potential surrogate is important for many reasons. In practice, information about surrogate strength will inform decisions regarding whether to measure the surrogate in a future study (especially if it is costly or invasive to measure) and/or whether to use the surrogate to assess treatment effectiveness in a future study. In addition, a number of novel statistical methods for using surrogates in future studies can only be applied when the surrogate is strong. For example, Parast \etal \cite{parast2019using} propose a robust nonparametric procedure to test for a treatment effect using surrogate marker information measured prior to the end of the study in a time‐to‐event outcome setting, but they rely on the assumption that the surrogate is sufficiently strong. As another example, Price \etal \cite{price2018estimation} propose constructing a function $\psi_a(\cdot)$ of the surrogates that leads to some optimality  properties, but it can be shown that this method should only be used with a strong surrogate. 

Specifically, they construct an optimal $\psi_a(\cdot)$ in terms of minimizing the following mean squared error (MSE)
\begin{equation}
\Msc_\psi = \Ebb\left[e(\bX) \left\{Y\sone - \psi_1(\bX, \bS\sone)\right\}^2 \right] + \Ebb\left[\{1-e(\bX)\} \left\{Y\szero - \psi_0(\bX, \bS\szero)\right]^2 \right], \qquad e(\bx) = P(A = 1 | \bX)\label{MSEPSI}
\end{equation}
under the constraint that it satisfies the so-called Prentice definition, i.e., 
\begin{align}\label{prentice}
    E(Y\sone - Y\szero) = 0 \text{ if and only if } E\left\{\psi_1(\bX, \bS\sone) - \psi_0(\bX, \bS\szero)\right\} = 0.
\end{align} 
The optimal transformations in this case are shown to be
$
\psi_a(\bx, \bs) = \Ebb\{Y\sa | \bX = \bx, \bS\sa = \bs\}.
$
In addition to this appealing optimality property, this proposal also resolves the so-called ``surrogate paradox." The paradox states that the treatment could have a positive causal effect on a univariate surrogate $S$ which could have a positive correlation with the outcome, but the treatment could yet have a negative effect on the outcome. The Price surrogate resolves the surrogate paradox by definition. The treatment effect on the surrogate $\Ebb(\Deltacheck)$ is by definition equivalent to the treatment effect on the outcome: 
$
    \Ebb(\Deltacheck) = \Ebb\{\psi_1(\bX, \bS\sone)\} - \Ebb\{\psi_0(\bX, \bS\szero)\} = \Delta.\label{psi-mu} 
$

However, the Price surrogate's resolution of the surrogate paradox is in some sense too good. The surrogate paradox is thus resolved for every potential set of surrogates, good and bad. Even if the surrogates $\bS$ are completely unrelated to the outcome $Y$, $\Ebb(\Deltacheck) = \Delta$. In fact, the power to detect a treatment effect in a future study using $\Deltacheck$ actually increases as the surrogate becomes weaker (explaining less of the treatment effect), if the Prentice definition does not hold. Consider the toy example depicted in Figure \ref{toy-eg} (see Appendix \ref{fig1-sim} for details of the data generation). 
\begin{figure}
\centering
\begin{tabular}{c}
\includegraphics[width =\textwidth]{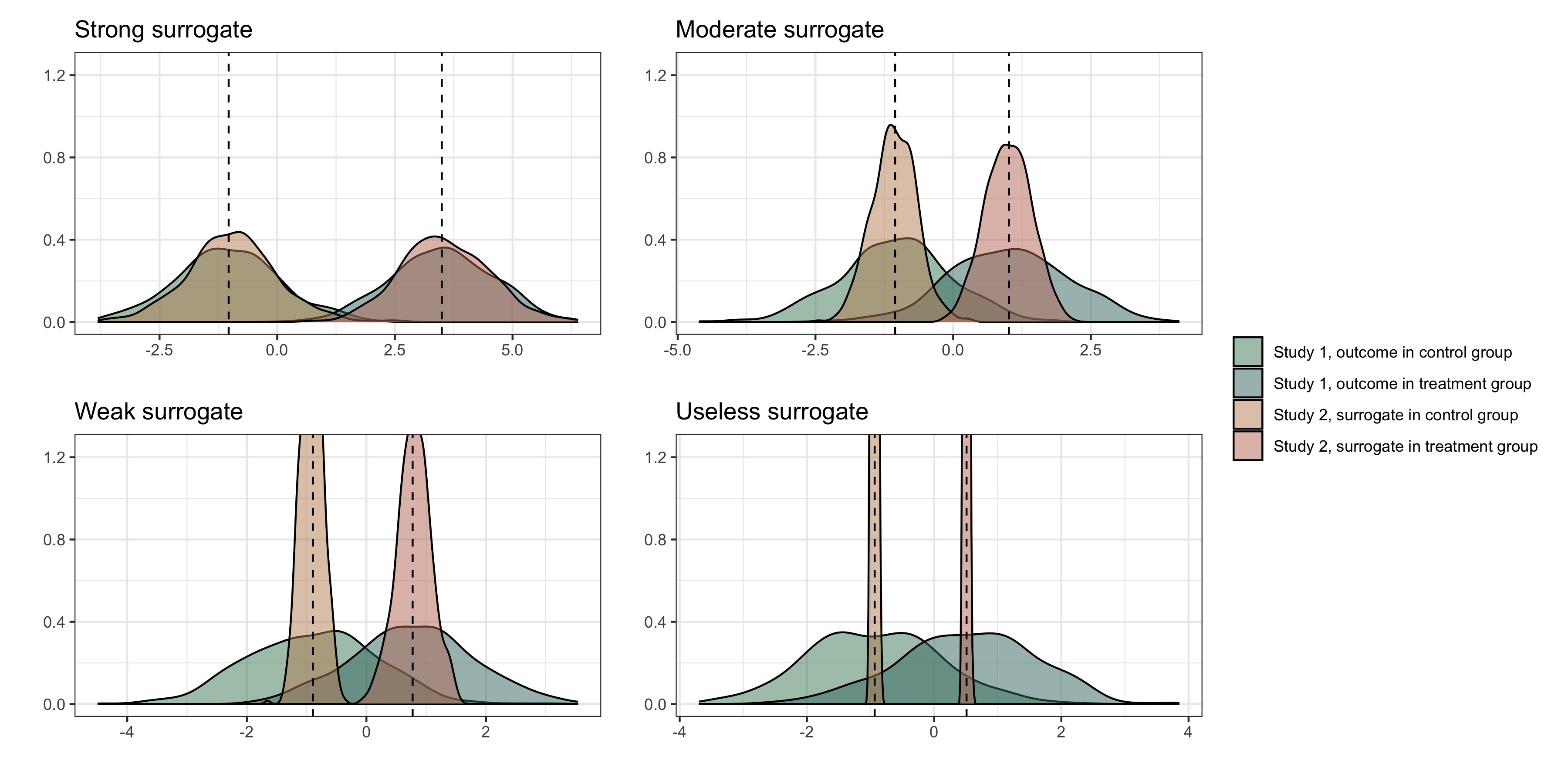}
\end{tabular}\vspace{0.2in}
\caption{Distribution of outcomes $Y$ in an initial study (Study 1) and surrogate functions (approximating the outcome) $\psi_1$ (surrogate in treatment group) and $\psi_0$ (surrogate in control group) in a future study (Study 2) based on simulated data. Vertical dotted lines correspond to the means of the treatment groups in the first study. See Appendix \ref{fig1-sim} for simulation details. Surrogate functions $\psi_0, \psi_1$ were estimated via a correctly specified linear regression in the first study and then applied to the second study population. When the surrogate was strong, the distribution of the surrogate in the second study was quite close to the distribution of the outcome in the first study. However, when the surrogate was weak, the distribution of the surrogate in the second study clustered around the group mean from the first study.}
\label{toy-eg}
\end{figure}
As the strength of the surrogate increases, $\psi_a(\cdot)$ becomes more like the true outcome $Y_i\sa$. As the surrogate becomes weaker, $\psi_a(\cdot)$ becomes more like $\Ybar_a$ the mean outcome in treatment group $A = a$ from the first study. This means that a weak surrogate ensures that the treatment effect in the second study will be identical to the treatment effect in the first study because the distribution of $\psi_a(\bXcheck_i, \bScheck_i)$ will cluster closely around the estimate of $E(Y\sa)$ from the first study (see specifically, the fourth panel of Figure \ref{fig1-sim}). In this scenario, the second study is providing no new information about the treatment. Thus, the Price surrogate should only be used with a strong $\bS$, and its use with a weak or possibly even moderately strong set of surrogates may not be advisable.

Therefore, for both decision-making purposes and to use statistical methods that take advantage of strong surrogates, methods are needed to rigorously quantify the strength of the proposed set of surrogates before using (functions of) the surrogates in practice. In the following section, we propose a model-free method for estimating the strength of a possibly high-dimensional set of surrogates in observational studies.

\subsection{Proposed method to evaluate surrogate strength}\label{sec:pte}
In this section we propose a general method for evaluating the usefulness of a set of surrogates. This can be used to evaluate the suitability of the Price surrogate but can also be used with other surrogate transformations, and it is appropriate for $\bS$ of any dimension. To evaluate the surrogates' usefulness, we follow \cite{parast2019assessing, parast2016robust, agniel2020evaluation} in evaluating the \textit{residual treatment effect}
\begin{align}
    \Delta_{\bS} = \Ebb\left\{\psi_1(\bX, \bS) - \psi_0(\bX, \bS)\right\},  \label{def:delta-s}
\end{align}
where $\psi_a(\bx, \bs) = E(Y\sa | \bX = \bx, \bS\sa = \bs)$, as defined above. This quantity corresponds to the treatment effect conditional on the distribution of $\bS\szero$ and $\bS\sone$ both being equal to the distribution of $\bS$ (all conditional on $\bX$). If $\bS\szero, \bS\sone$ are independent of $Y\szero, Y\sone$ conditional on $\bX$, then $\Delta_{\bS} = \Delta$. In contrast, if all of the treatment effect can be attributed to $\bS$, then $\Delta_{\bS}$ should be close to 0. We may use the residual treatment effect to quantify \textit{how much} of the treatment effect the surrogates account for via the \textit{proportion of treatment effect explained} (PTE)
\begin{align}
    R_{\bS} = \frac{\Delta - \Delta_{\bS}}{\Delta} = 1 - \frac{\Delta_{\bS}}{\Delta}.
\end{align}

The quantity $\Delta_{\bS}$ is defined such that it aims to quantify the effect of the treatment on the outcome that is \textit{not} captured by the effect of the treatment on the surrogate, by forcing the conditional distribution of the potential surrogate in both treatment groups to be the same. Based on this definition, $\Delta - \Delta_{\bS}$ then parallels the natural indirect effect quantity \cite{joffe2009related,vanderweele2013surrogate} which is often used to assess the proportion mediated in a mediation analysis. However, there is an important distinction between assessing a variable (or variables) for surrogacy versus mediation. As described by \cite{vanderweele2013surrogate}, the aim of identifying a mediator is determining whether the effect of treatment operates through the mediator itself e.g., through some biological pathway. Often, a good surrogate marker is similarly conceptualized as a variable through which the treatment operates, but this is not necessarily required; a variable can be a good surrogate if it captures the treatment effect on the outcome, even if the treatment effect does not operate through the variable itself. 

\begin{remark} While $R_S$ is easy to understand and interpret, and it has analogues in the mediation literature, it is not guaranteed to lie between 0 and 1. We give conditions in Section \ref{pte-assump} that guarantee that $R_S \in (0,1)$. In situations when these conditions are not or are unlikely to be met, one may desire a secondary definition of the PTE that can be estimated in all situations similar to \cite{wang2020model}. Conceptually, here we would consider a good surrogate to be one whereby the surrogate transformation substantially improves the MSE for predicting $Y\sa$ over a \textit{null} transformation that does not use $\bS$. The MSE of the null transformation would be defined as:
\begin{align*}
    \Msc_0 &= \Ebb\left(e(\bX) \left[Y\sone - m_1\{\bX\}\right]^2 \right) + \Ebb\left(\{1-e(\bX)\} \left[Y\szero - m_0\{\bX\}\right]^2 \right)
\end{align*}
where $m_a(\bx) = \Ebb(Y\sa |\bX = \bx).$
The secondary MSE-based definition of PTE could then be defined as the proportion of the null MSE that is reduced by controlling for the surrogates:
\begin{align}
    R_M = \frac{\Msc_0 - \Msc_\psi}{\Msc_0} = 1 - \Msc_\psi/\Msc_0
\end{align}
where $\Msc_\psi$ is defined in \eqref{MSEPSI}.
\end{remark}

\section{Assumptions and Estimation}

\subsection{Identifying assumptions}\label{assump-ident}
We make the three typical assumptions of treatment effect estimation: consistency, positivity, and no unmeasured confounding. Specifically, we assume that the observed values of $\bS$ and $Y$ when $A = a$ are identical to the counterfactuals $\bS\sa$ and $Y\sa$ such that
$\bS = \bS\sone A + \bS\szero(1 - A)$ and $Y = Y\sone A + Y\szero(1 - A).$
We furthermore assume that $\bX$ contains all confounders of the effects of $A$ on the surrogates and the outcome, such that the treatment $A$ is as good as randomized conditional on the covariates $\bX$
\begin{align}
\left.\{\bS\szero, \bS\sone, Y\szero, Y\sone\} \independent A \right| \bX.\label{psr}
\end{align}
In addition, we assume two forms of positivity, which ensures that individuals in the two study arms are not too different from one another. First, the usual positive probability of receiving either treatment for some $\epsilon_1 > 0$
\[
P\{\epsilon_1 < e(\bX) < 1-\epsilon_1\} = 1,
\]
and a related assumption that further conditions on the surrogates
\[
P\{\epsilon_2 < \pi(\bX, \bS) < 1-\epsilon_2\} = 1,
\]
for some $\epsilon_2 > 0$ where $\pi(\bS, \bX) = P(A = 1 | \bS, \bX)$ is the \textit{surrogate score} -- akin to the propensity score \cite{rosenbaum1983central} -- which has been proposed for a different purpose in \cite{Athey2016}. Notably, because $\pi(\bx, \bs) = \frac{P(\bS = \bs | \bX = \bx, A = 1)e(\bx)}{P(\bS = \bs | \bX = \bx)} = \frac{P(\bS\sone = \bs | \bX = \bx)e(\bx)}{P(\bS\sone = \bs | \bX = \bx)e(\bx) + P(\bS\szero = \bs | \bX = \bx)\{1-e(\bx)\}}$, these two positivity conditions ensure that the conditional distribution of the counterfactual surrogates under treatment and control cannot be too different from one another, i.e., ensuring overlap.

We additionally give regularity conditions facilitating estimation in Appendix \ref{assump-reg}.

\subsection{Required assumptions to ensure interpretation of $R_\bS$ as a proportion}\label{pte-assump}

The interpretation of $R_\bS$ as the PTE depends on it actually being a proportion, lying between 0 and 1. We adapt conditions in \cite{wang2002measure} and \cite{agniel2020evaluation} which ensure that $0 \leq R_\bS \leq 1$. These conditions are as follows (assuming without loss of generality that $\Delta > 0$):
\begin{align}
\int \psi_1(\bx, \bs)d\left\{F_{\bX, \bS\sone}(\bx, \bs) - F_{\bX, \bS}(\bx, \bs)\right\} \geq \int \psi_0(\bx, \bs)d\left\{F_{\bX, \bS\szero}(\bx, \bs) - F_{\bX, \bS}(\bx, \bs)\right\} \text{ and }\label{as1}
\end{align}
\begin{align}
\int \left\{\psi_1(\bx, \bs) - \psi_0(\bx, \bs)\right\}dF_{\bX, \bS}(\bx, \bs) \geq 0.\label{as2}
\end{align}
If $\Delta < 0$, then one would require \eqref{as1} and \eqref{as2} to hold with inequalities reversed. 

Because $F_{\bX, \bS}$ is a mixture distribution $F_{\bX, \bS}(\bx, \bs) = e(\bx) F_{\bX, \bS\sone}(\bx, \bs) + \{1-e(\bx)\}F_{\bX, \bS\szero}(\bx, \bs)$, the first condition \eqref{as1} may be rewritten:
\begin{align}
\int \left\{e(\bx)\psi_1(\bx, \bs) + \{1-e(\bx)\}\psi_0(\bx, \bs)\right\}d\left\{F_{\bX, \bS\sone}(\bx, \bs) - F_{\bX, \bS\szero}(\bx, \bs)\right\} \geq 0.\label{as1star}
\end{align}
The condition \eqref{as2} ensures that $\Delta_S \geq 0$, i.e., that the residual treatment effect is in the same direction as the overall treatment effect $\Delta$. Condition \eqref{as1star}, which ensures that $\Delta \geq \Delta_S$, requires that, roughly speaking, a propensity-weighted mixture of the two conditional mean functions $e(\bx)\psi_1(\bx, \bs) + \{1-e(\bx)\}\psi_0(\bx, \bs)$ is larger when $\bs$ takes values from the distribution of the counterfactual surrogates under treatment than if it took values from the distribution under control.  These two conditions are analogues of conditions (A6) and (A7) in \cite{agniel2020evaluation}.

\subsection{Identification}

In this section, we show how $\Delta_{\bS}$ may be characterized via functionals of the observed data distribution. Specifically, 
\begin{align}
    \Delta_{\bS} &= \Ebb\left\{\mu_1(\bX, \bS) - \mu_0(\bX, \bS)\right\}\label{om}\\
    &= \Ebb\left[\frac{AY}{\pi(\bX, \bS)} - \frac{(1-A)Y}{\{1-\pi(\bX, \bS)\}}\right]\label{ipw}\\
    &= \Ebb\left[\frac{AY - \{A - \pi(\bX, \bS)\}\mu_1(\bX, \bS)}{\pi(\bX, \bS)} - \frac{(1-A)Y + (A - \pi(\bX, \bS))\mu_0(\bX, \bS)}{1-\pi(\bX, \bS)}\right],\label{dr}
\end{align} 
where $\mu_a(\bx, \bs) = \Ebb(Y | \bX = \bx, A = a, \bS = \bs)$. The result \eqref{om} follows from the definition \eqref{def:delta-s} and the fact that $\psi_a(\bX, \bS\sa) = \mu_a(\bX, \bS)$ because of \eqref{psr}, and the other results follow shortly thereafter, following familiar paths as arguments for identification of the average treatment effect in other contexts. 
These results show that the residual treatment effect may be identified without knowledge of the mean function for the outcome via \eqref{ipw}, and gives an augmented inverse probability weighting \cite{bang2005doubly} version of the estimand \eqref{dr} .$\Delta$ may also be identified using similar functionals with $\mu_a(\bX, \bS)$ replaced by $m_a(\bX)$ and $\pi(\bX, \bS)$ replaced by $e(\bX)$:
\begin{align}
    \Delta &=  \Ebb\left[\frac{AY - \{A - e(\bX)\}m_1(\bX)}{e(\bX)} - \frac{(1-A)Y + (A - e(\bX))m_0(\bX)}{1-e(\bX)}\right]\label{drdelta}
\end{align}
where $m_a(\bx) = E(Y | \bX = \bx, A = a)$. In Section \ref{est-sec}, we propose an estimation approach using the expressions in \eqref{dr} and \eqref{drdelta}.

\begin{remark}
If one wished to pursue the alternative PTE definition using $\Msc_\psi$ and $\Msc_0$, these quantities may also be identified from the observed data. Let $\eta^2_a(\bx, \bs) = \Ebb\{Y - \mu_a(\bx, \bs)\}^2$ be the conditional (on $\bX$ and $\bS$) variance of $Y$ in treatment arm $A = a$. Further define $\sigma_a^2(\bx) = \Ebb\left\{\eta^2_a(\bx, \bS) | \bX = \bx, A = a\right\}$ to be the expectation of $\eta^2_a(\bx, \bS)$ over the conditional distribution of $\bS$ given $A = a, \bX = \bx$. Then using a similar argument as above, we have that 
\begin{align*}
    \Msc_\psi &= \Ebb\left[e(\bX)\sigma^2_1(\bX) + \{1-e(\bX)\}\sigma^2_0(\bX)\right]\\ &= \Ebb\left(A\eta_1(\bX, \bS) + (1-A)\eta_0(\bX, \bS) \right) \\
    &= \Ebb\left(A\eta_1(\bX, \bS) - \{A - e(\bX)\}\sigma^2_1(\bX) + (1-A)\eta_0(\bX, \bS) + (A - e(\bX))\sigma^2_0(\bX)\right)
\end{align*}
with similar expressions holding for $\Msc_0$ \textit{mutatis mutandis}. The first equality is analogous to an outcome-model-type identification, as in \eqref{om}, the second equality is an inverse-probability-type identification, as in \eqref{ipw}, and the third is a doubly-robust-type identification, as in \eqref{dr}. One would need to modify slightly the cross-fitting procedure outlined in the next section to incorporate $\sigma_a^2(\bx)$, which would require two disjoint folds of data for estimation: one to estimate $\eta_a^2(\bx, \bs)$ and a second to estimate $\sigma_a^2(\bx)$.
\end{remark} 

\subsection{Estimation \label{est-sec}}
We propose an estimation approach utilizing the augmented inverse probability weighting versions of the estimands in \eqref{dr} and \eqref{drdelta}. The traditional form of these estimators would take the form (for $\Delta_S$)
\begin{align*}
    \Deltahat_\bS(\Osc, \pihat, \muhat_0, \muhat_1) = n\inv \sumin \frac{A_iY_i - \{A_i - \pihat(\bX_i, \bS_i)\}\muhat_1(\bX_i, \bS_i)}{\pihat(\bX_i, \bS_i)} - n\inv \sumin \frac{(1-A_i)Y_i + \{A_i - \pihat(\bX_i, \bS_i)\}\muhat_0(\bX_i, \bS_i)}{1-\pihat(\bX_i, \bS_i)}
\end{align*}
with $\muhat_a(\bx, \bs) = \Ehat\{Y | \bX = \bx, A = a, \bS = \bs\}$ an estimate of the conditional outcome function and $\pihat(\bx, \bs) = \Phat(A = 1 | \bS = \bs, \bX = \bx)$, and (for $\Delta$)
\begin{align*}
    \Deltahat(\Osc, \ehat, \mhat_0, \mhat_1) = n\inv \sumin \frac{A_iY_i - \{A_i - \ehat(\bX_i)\}\mhat_1(\bX_i)}{\ehat(\bX_i)} - n\inv \sumin \frac{(1-A_i)Y_i + \{A_i - \ehat(\bX_i)\}\mhat_0(\bX_i)}{1-\ehat(\bX_i)}
\end{align*}
with $\mhat_a(\bx) = \Ehat\{Y | \bX = \bx, A = a\}$  and $\ehat(\bx) = \Phat(A = 1 | \bX = \bx)$ estimates of the corresponding mean functions and propensity score.
We discuss the doubly robust property further in Section \ref{discussion}.

 Here, we aim to specify our approach as generally as possible, such that in practice, one could choose any reasonable approach to obtaining estimates for $\muhat_a(\bx, \bs)$,  $\pihat(\bx, \bs)$, $\mhat_a(\bx)$, and $\ehat(\bx)$ while maintaining the ability to make inference about the PTE $R_\bS$. Certainly, if the dimensions of $\bX$ and $\bS$ are low, one could use simple parametric functions or, if very low (e.g., $X, S$ both scalar random variables), one could use nonparametric kernel smoothing. As we are specifically interested in settings where $\bS$, and possibly even $\bX$, may be high-dimensional, we do not restrict ourselves to settings where the above may be feasible, and we allow for machine learning and high-dimensional regression approaches to be used for estimating these functions. In our simulations, we specifically examine two versions of our proposed estimation approach: one that uses the Super Learner \cite{van2007super} and one that uses the relaxed lasso \cite{meinshausen2007relaxed} to estimate each of the quantities: $\muhat_a(\bx, \bs)$, $\pihat(\bx, \bs)$, $\mhat_a(\bx)$, and $\ehat(\bx)$. Details are described in Section \ref{sims-sec}.

Importantly, in these settings when the dimensions of $\bX$ and $\bS$ may be high and/or when flexible machine learning models are used to estimate the quantities $\{\muhat_0(\cdot), \muhat_1(\cdot), \pihat(\cdot), \mhat_0(\cdot), \mhat_1(\cdot)$, $\ehat(\cdot)\}$, additional steps that preclude overfitting and facilitate asymptotic analysis should be considered. We propose to use the \textit{cross-fitting} approach of \cite{Chernozhukov2017}, where (for a given positive integer $K > 1$), the observed data are randomly partitioned into $K$ sets $(\Osc_k)_{k=1,...,K}, \bigcup_k \Osc_k = \Osc, \bigcap_k \Osc_k = \emptyset$, with the indices corresponding to $\Osc_k$ denoted $I_k$, such that each set $\Osc_k$ has $m = |I_k| = n/K$ observations in it (assuming $n$ is a multiple of $K$). The estimator for $\Delta_\bS$ then takes the form
\begin{align}
    \Deltahat_\bS = K\inv\sum_{k=1}^K \Deltahat_{\bS,k} = K\inv\sum_{k=1}^K \Deltahat_{\bS}(\Osc_k, \pihat_{-k}, \muhat_{0-k}, \muhat_{1-k})
\end{align}
where
\begin{align*}
    \Deltahat_{\bS, k} &= m\inv \sum_{i \in I_k} \frac{A_iY_i - \{A_i - \pihat_{-k}(\bX_i, \bS_i)\}\muhat_{1-k}(\bX_i, \bS_i)}{\pihat_{-k}(\bX_i, \bS_i)} - \\
    &\qquad m\inv \sum_{i \in I_k} \frac{(1-A_i)Y_i + \{A_i - \pihat_{-k}(\bX_i, \bS_i)\}\muhat_{0-k}(\bX_i, \bS_i)}{1-\pihat_{-k}(\bX_i, \bS_i)}
\end{align*}
and $(\pihat_{-k}, \muhat_{0-k}, \muhat_{1-k})$ are estimated on all of the data excluding $\Osc_k$: $\Osc_k^c = \Osc / \Osc_k$. 

The estimator for $\Delta$ uses a similar cross-fitting procedure:
\begin{align}
\Deltahat &= K\inv\sum_{k=1}^K \Deltahat_{k} = K\inv\sum_{k=1}^K \Deltahat(\Osc_k, \ehat_{-k}, \mhat_{0-k}, \mhat_{1-k})\\
\Deltahat_{k} &= m\inv \sum_{i \in I_k} \frac{A_iY_i - \{A_i - \ehat_{-k}(\bX_i)\}\mhat_{1-k}(\bX_i)}{\ehat_{-k}(\bX_i)} - m\inv \sum_{i \in I_k} \frac{(1-A_i)Y_i + \{A_i - \ehat_{-k}(\bX_i)\}\mhat_{0-k}(\bX_i)}{1-\ehat_{-k}(\bX_i)}
\end{align}
where $\mhat_{a-k}(\bx) = \Ehat(Y | A = a, \bX = \bx)$ is the outcome model conditional only on the pre-treatment covariates $\bX$ and $\ehat_{-k}(\bx) = \Phat(A = 1 | \bX = \bx)$ is the estimated propensity score, both estimated on $\Osc_k^c$. Finally, the PTE may be estimated as
\[
\Rhat_{\bS} = \frac{\Deltahat - \Deltahat_{\bS}}{\Deltahat} = 1 - \frac{\Deltahat_{\bS}}{\Deltahat}.
\]



\section{Asymptotic Behavior, Inference, and Variance Estimation}\label{asymp}
In this section, we show that our approach (and specifically, the use of cross-fitting) yields a convenient asymptotic distribution for the PTE estimator (given certain regularity conditions we outline), and we demonstrate how this may be used for making inference on $\Rhat$. 

The key conditions are on the rate of convergence of the nuisance functions $\{\muhat_0(\cdot), \muhat_1(\cdot), \pihat(\cdot), \mhat_0(\cdot), \mhat_1(\cdot)$, $\ehat(\cdot)\}$. These rates of convergence are allowed to be quite slow. Specifically, each estimated nuisance function is allowed to converge to the true nuisance function at a rate as slow as $n^{-1/4}$. Furthermore, because of sample-splitting, the form of the asymptotic variance and its plug-in estimate are convenient to use. We now state the main result more formally
\begin{proposition}
Let
\begin{align}
    \phi_1(\bO; \Delta, \eta_1) = \frac{AY - \{A - e(\bX)\}m_1(\bX)}{e(\bX)} - \frac{(1-A)Y + \{A - e(\bX)\}m_0(\bX)}{1-e(\bX)} - \Delta,
\end{align}
and
\begin{align}
    \phi_2(\bO; \Delta_\bS, \pi, \mu_0, \mu_1) = \frac{AY - \{A - \pi(\bX, \bS)\}\mu_1(\bX, \bS)}{\pi(\bX, \bS)} - \frac{(1-A)Y + \{A - \pi(\bX, \bS)\}\mu_0(\bX, \bS)}{1-\pi(\bX, \bS)} - \Delta_\bS.
\end{align}
If, in addition to the assumptions in Section \ref{assump-ident} and Appendix \ref{assump-reg}, we also have, for a sequence of positive constants $\delta_n, n= 1, ..., \infty$ that 
\begin{align*}
    &E\left[\{\muhat_{a-k}(\bX, \bS) - \mu_a(\bX, \bS)\}^2\right] \times E\left[\{\pihat_{-k}(\bX, \bS) - \pi(\bX, \bS)\}^2\right] \leq \delta_n \nnhalf,\\
    &E\left[\{\muhat_{a-k}(\bX, \bS) - \mu_a(\bX, \bS)\}^2\right] + E\left[\{\pihat_{-k}(\bX, \bS) - \pi(\bX, \bS)\}^2\right] \leq \delta_n, \\
    &P\{\epsilon_2 < \pihat_{-k}(\bS, \bX) < 1-\epsilon_2\} = 1
\end{align*}
and 
\begin{align*}
    &E\left[\{\mhat_{a-k}(\bX) - m_a(\bX)\}^2\right] \times E\left[\{\ehat_{-k}(\bX) - e(\bX)\}^2\right] \leq \delta_n \nnhalf,\\
    &E\left[\{\mhat_{a-k}(\bX) - m_a(\bX)\}^2\right] + E\left[\{\ehat_{-k}(\bX) - e(\bX)\}^2\right] \leq \delta_n,\\
    &P\{\epsilon_1 < \ehat_{-k}(\bX) < 1-\epsilon_1\} = 1,
\end{align*}
then
\begin{align*}
    &\sqrt{n}\sigma^{-1}(\Rhat_\bS - R_{\bS}) \rightarrow N(0, 1), \quad \mbox{where}\\
    &\sigma^2 = \Delta^{-2}E\left\{\phi_1(\bO, \Delta, e, m_0, m_1)^2\right\} + \Delta_{\bS}\Delta^{-4}E\left\{\phi_2(\bO, \Delta_\bS, \pi, \mu_0, \mu_1)^2\right\} -\\
    &\qquad 2\Delta_{\bS}\Delta^{-3}E\left\{\phi_1(\bO, \Delta, e, m_0, m_1)\phi_2(\bO, \Delta_\bS, \pi, \mu_0, \mu_1)\right\}.
\end{align*}
\end{proposition}
The consequence of this result is that a straightforward $(1-\alpha)\%$ confidence interval for $R_\bS$ can be obtained as 
\begin{align}\label{asymp-ci}
    \Csc = (\Rhat_\bS - z_{1-\alpha/2}\sigmahat, \Rhat_\bS + z_{1-\alpha/2}\sigmahat),
\end{align} with $z_q$ the $q$th quantile of the standard normal distribution and
\begin{align*}
    &\sigmahat^2 = n\inv\sumin\left(\Deltahat^{-2}\phihat_{1i}^2 + \Deltahat_{\bS}^2\Deltahat^{-4}\phihat_{2i}^2 - 2\Deltahat_{\bS}\Deltahat^{-3}\phihat_{1i}\phihat_{2i}\right).
\end{align*}
where 
\[\phihat_{1i} = \phi_1(\bO_i, \Deltahat, \ehat_{-k_i}, \mhat_{0-k_i}, \mhat_{1-k_i}), \phihat_{2i}=\phi_2(\bO_i, \Deltahat_\bS, \pihat_{-k_i}, \muhat_{0-k_i}, \muhat_{1-k_i})\}\trans
\] and $k_i = I\{k : i \in I_k\}$. In addition, it follows that $P(R_\bS \in \Csc) \rightarrow 1- \alpha$ because, under these assumptions, based on the arguments in \cite{Chernozhukov2017},
\begin{align}
    \sqrt{n}\Sigma^{-\frac{1}{2}}(\bthetahat - \btheta)  \rightarrow N(\bzero, I)
\end{align}
where $\bthetahat = (\Deltahat, \Deltahat_\bS)\trans, \btheta = (\Delta, \Delta_\bS)\trans, \Sigma = \Ebb(\bphi\bphi\trans)$ and this holds when replacing $\Sigma$ by $\Sigmahat = n\inv\sumin\bphi_i\bphi_i\trans$, $\bphihat_i = (\phi_{1i}, \phi_{2i})\trans$.

\section{Simulations \label{sims-sec}}

\subsection{Simulation Overview}
We used two approaches to estimate the six functions $\pi(\bX, \bS), \mu_1(\bX, \bS), \mu_0(\bX, \bS), \pi(\bX), m_1(\bX), m_0(\bX)$, yielding two different versions of our proposed estimator. In one, we used the Super Learner \cite{van2007super}, which finds an optimal combination of a set of candidate models or learners, as implemented in the \texttt{SuperLearner} package \cite{polley2019package} in \texttt{R}. We denote this estimator ``DR-SL" in what follows. In the second version, which we call "DR-lasso" we use the relaxed lasso \cite{meinshausen2007relaxed}, as implemented in the \texttt{glmnet} package \cite{glmnet}, to estimate all needed functions. For all simulations, we truncated estimates of the propensity and surrogate scores so that $0.01 \leq \pihat_{-k}(\bX_i, \bS_i), \ehat_{-k}(\bX_i) \leq 0.99$ for all $i$.

It is our understanding that there are no currently available methods to estimate the PTE of a high-dimensional surrogate. However, since there are available methods to measure the strength of high-dimensional mediators, we compare our proposed approach to these available methods. While the goals of mediation analysis and surrogate markers analysis are different and the assumptions necessary differ, this allows us to offer some reasonable comparison to methodology that is currently available, rather than no comparison at all. Thus, to fairly compare, we purposefully set up our simulations such that $\bS$ is a mediator in that it lies on the causal pathway between $A$ and $Y$. While mediation methods are attempting to estimate a distinct quantity from the PTE we are interested in, in the following simulations, the estimation of the ``proportion mediated" (which is used in mediation) and the PTE that we are interested in are, in fact, the same. 

Specifically, in our simulations we compare our proposed approach to three methods for high-dimensional mediation analysis: HIMA (High-dimensional mediation analysis, \cite{zhang2016estimating}) as implemented in the \texttt{HIMA} package \cite{hima}; BAMA (Bayesian mediation analysis, \cite{song2018bayesian}) as implemented in the \texttt{bama} package \cite{bama}; and HILMA (High-dimensional linear mediation analysis, \cite{zhou2020estimation}) as implemented in the \texttt{freebird} package. All three of these methods propose $p+1$ linear models: $p$ models for the surrogates/mediators
\[
S_{ij} = \alpha_{0j} + \alpha_jA_i + \sum_{k=1}^q\bgamma^*_{jk}X_{ik} + e_{ij}, j = 1,...,p; i = 1, ..., n,
\]
and one outcome model
\[
Y_i = \beta_0 + \Delta_S A_i + \sum_{j=1}^p\beta_jS_{ij} + \sum_{k=1}^q\gamma_kX_{ik} + \epsilon_i,
\]
though the implementation of freebird does not allow for the inclusion of covariates $X_{ik}$.
Using these models, the overall treatment effect can be identified as $\Delta = \Delta_S + \sum_{j=1}^p\alpha_j\beta_j$, and as above the PTE (or proportion mediated) may be identified as $R = 1 - \Delta_S/\Delta.$ The HIMA procedure employs sure independence screening and penalized least squares with the minimax concave penalty \cite{zhang2010nearly} to set some values of $\alpha_j\beta_j$ to be exactly 0. Similarly, BAMA employs Bayesian shrinkage estimation to estimate all parameters (including some at exactly 0) under normality assumptions on the error terms. Finally, the freebird approach debiases a pilot scaled lasso estimator. 

We computed 95\% confidence intervals (CIs) for $R$ from the results given in Section \ref{asymp} for the two proposed approaches. We computed 95\% credible intervals for the BAMA estimator from the 2.5\% and 97.5\% quantiles of the posterior distribution of $1 - \frac{\Delta_S}{\Delta_S + \sum_{j=1}^p\alpha_j\beta_j}.$ CIs were computed for the HIMA and freebird implementations because they are not available. 

\subsection{Simulation Setup}
We constructed two sets of simulations for a total of 10 simulation settings to assess the performance of our proposed approach in both low- and high-dimensional settings. 

For the first set of simulations, the data-generating mechanism for $\bS\sa$ was linear in $\bX$, the data-generating mechanism for $Y\sa$ was linear in $\bX$ and $\bS\sa$, and the propensity score was linear on the log-odds scale. Given this data-generating mechanism, we would expect that all methods (proposed and comparisons) should perform reasonably well. We set the dimension of $\bS$ ($p$) and of $\bX$ ($q$) to be 100. We let  $X_{ij} \sim N(0, 1), A_i \sim \text{Bernoulli}\{\pi(\bX_i)\}, \pi(\bX_i) = \text{logit}(\bgamma\trans\bX_i)$, where $\bgamma \sim N(0, 1)$. The surrogates were generated as $\bS\sa_i = \balpha_a + \bbeta_a\trans\bX_i + \be_i$, where $\alpha_{11} = 0.75, \alpha_{12} = 0.25, \alpha_{01} = \alpha_{02} = 0, \alpha_{1j} \sim U(0,1), \alpha_{0j} \sim U(-0.5, 0.5), j = 3, ... p$. And $(\beta_{11}, \beta_{12}, \beta_{13}, \beta_{14}, \beta_{15}) = (-1, -0.5, 0, 0.5, 1), (\beta_{01}, \beta_{02}, \beta_{03}, \beta_{04}, \beta_{05}) = (-2, -1.5, -1, -0.5, 0), \beta_{aj} = 0, a=0,1, j > 5$. The outcome counterfactuals were given by $Y_i\sa = a\Delta_S + \sum_{j = 1}^{25} X_{ij} + S_{i1}\sa + S_{i2}\sa + \epsilon_i$. The errors $e_{ij}$ and $\epsilon_i$ were $N(0, \sigma^2)$. We let the sample size vary between $n = 100$ and $n = 500$ and the level of noise between $\sigma = 0.1$ and $\sigma = 0.5$, and we set $R = 0.5$. 

In the second set of simulations, the data-generating mechanism was less linear, including interactions between covariates in both the propensity score and the model for $\bS$, though the outcome model was still a simple linear combination of $\bX$ and $\bS$. This set of simulations should mimic what may happen in practice since all models are typically subject to some amount of mis-specification. We expected the nonlinearity to induce bias in the competing methods (which require linear models to hold). There were two pre-treatment covariates $X_1 \sim \text{Uniform}(-2, 5)$ and $X_2 \sim \text{Bernoulli}(0.5)$. The probability of treatment was determined by $P(A = 1 | \bX) = \text{logit}(-X1 + 2X_1X_2)$. The surrogates under treatment were given by $\bS\sone = 1.5 + (X_1 + X_2)\Isc_{10} + \be$ and under control by $\bS\szero = 2 + X_2\Isc_{10} - X_1X_2 + \be$ where $\Isc_{10}$ is a $p \times p$ matrix of 0s except it has 1s on the first 10 diagonal elements and $\be \sim N(0, \Sigma)$ and $\Sigma = 0.4\bone\bone\trans + 0.6I_p$ is a covariance matrix with common covariance $0.4$. Finally, the outcome counterfactuals were given by $Y\sa = a\Delta_S + X_1 + X_2 + \sum_{j = 1}^{15}S_j\sa + \epsilon, \epsilon \sim N(0, 1)$. We set the sample size to be $n=500$, and we varied the number of surrogates ($p = 50, 100, 500$), and the PTE (surrogates were either weak ($R = 0.41$) or strong ($R = 0.85$)).

\subsection{Results for first data-generating mechanism}
\begin{figure}
\centering
\begin{tabular}{c}
\includegraphics[width =\textwidth]{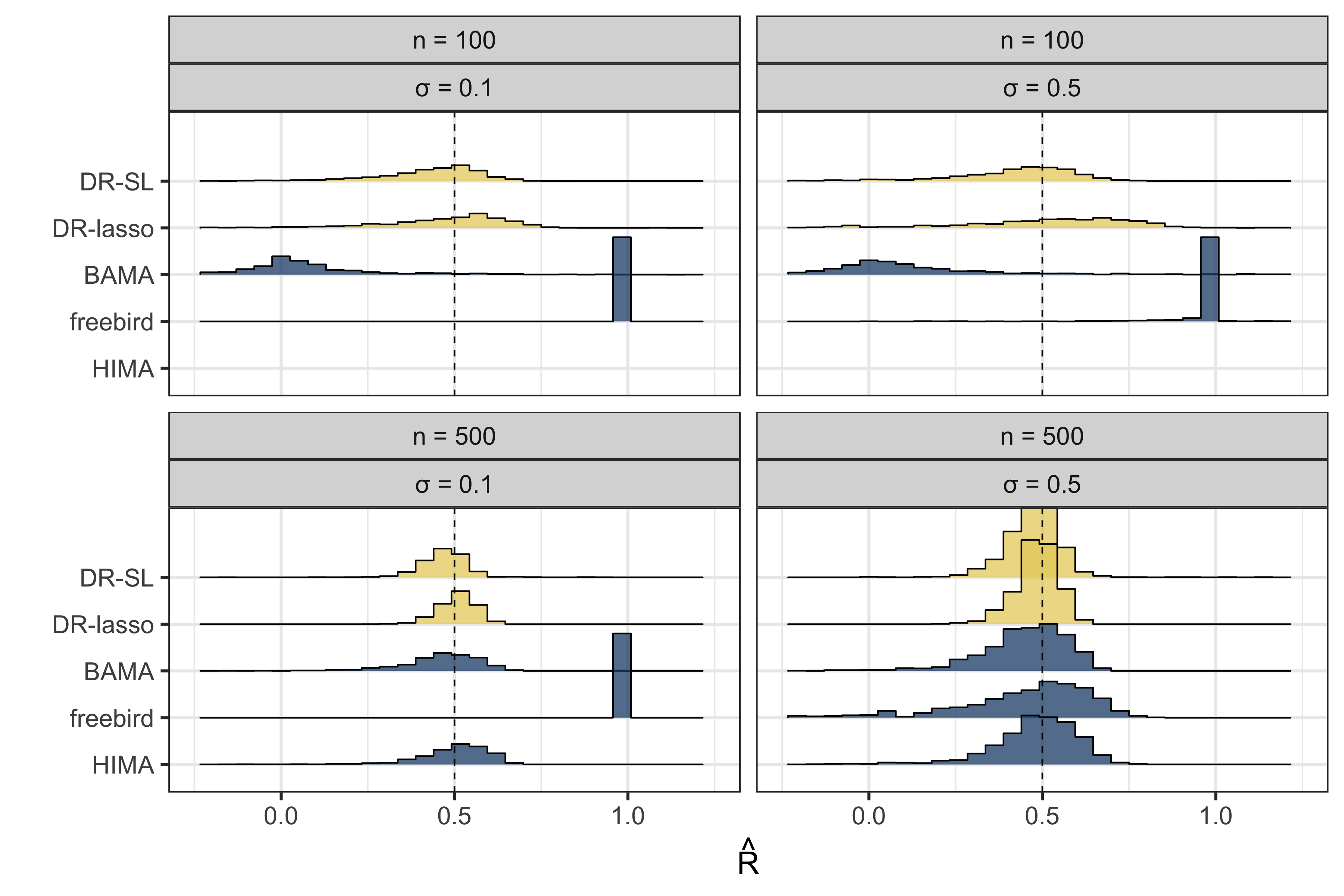}
\end{tabular}\vspace{0.2in}
\caption{Distribution of estimates of $\Rhat$ in first data-generating mechanism. Lighter shaded regions represent the distribution of the proposed estimators (``DR-SL" and ``DR-lasso"); darker shaded regions represent the distribution of the comparison estimators (``BAMA", ``freebird", and ``HIMA"). The true value of $R$ is given as a vertical dotted line at 0.5.}
\label{r-pl1}
\end{figure}

As expected, this data-generating mechanism was well approximated using linear models. When the sample size was large ($n = 500$) and $\sigma = 0.5$, all methods performed well, with relatively small bias (see Figure \ref{r-pl1}). The median $\Rhat$ estimates were 0.46 (BAMA), 0.48 (freebird), 0.49 (HIMA), 0.48 (DR-SL), and 0.49 (DR-lasso), all of which compare favorably to the true $R$ value of 0.5. The distribution of estimates for the proposed approaches was a bit tighter than for the competing methods. The empirical 2.5\% and 97.5\% quantiles were 0.30 and 0.62 for DR-SL, 0.35 and 0.58 for DR-lasso, 0.15 and 0.67 for HIMA, 0.11 and 0.63 for BAMA, and -0.85 and 0.71 for freebird. Results were similar for the sample size with less noise ($\sigma = 0.1$), except for the freebird approach which began to estimate the PTE to be exactly 1 in almost all simulations: the empirical 2.5\% and 97.5\% quantiles of the $\Rhat$ distribution for freebird were exactly 1. At the lower sample size ($n = 100$), HIMA did not run, while BAMA estimates tended to be near 0 and freebird estimates were again clustered tightly around 1. Our proposed approaches had median $\Rhat$ values of 0.46 (DR-SL) and 0.51 (DR-lasso), though with more variability than when $n = 500$. The median absolute deviation ($|\Rhat - R|$) was smaller for both DR-SL and DR-lasso than any of the competing approaches for all settings.

CI coverage for the proposed estimators tended to be conservative in all settings: coverage for DR-SL was 100\% ($n = 100, \sigma = 0.1$ and $n = 100, \sigma = 0.5$), 98\% ($n = 500, \sigma = 0.1$), and 99\% ($n = 500, \sigma = 0.5$), and for DR-lasso it was 100\% in all settings. 
On the other hand, BAMA CIs only had nominal coverage at lower sample sizes: 95\% ($n = 100, \sigma = 0.1$), 98\% ($n = 100, \sigma = 0.5$), 88\% ($n = 100, \sigma = 0.1$), and 84\% ($n = 500, \sigma = 0.5$). Even when BAMA CIs had nominal coverage, they were about four times larger than the CIs for the proposed estimators: BAMA CI half-lengths were 1.38 ($n = 100, \sigma = 0.1$) and 1.64 ($n = 100, \sigma = 0.5$), while the corresponding half-lengths were 0.47 and 0.57 for DR-SL and 0.57 and 1.02 for DR-lasso.

\subsection{Results for second data-generating mechanism}
\begin{figure}
\centering
\begin{tabular}{c}
\includegraphics[width =\textwidth]{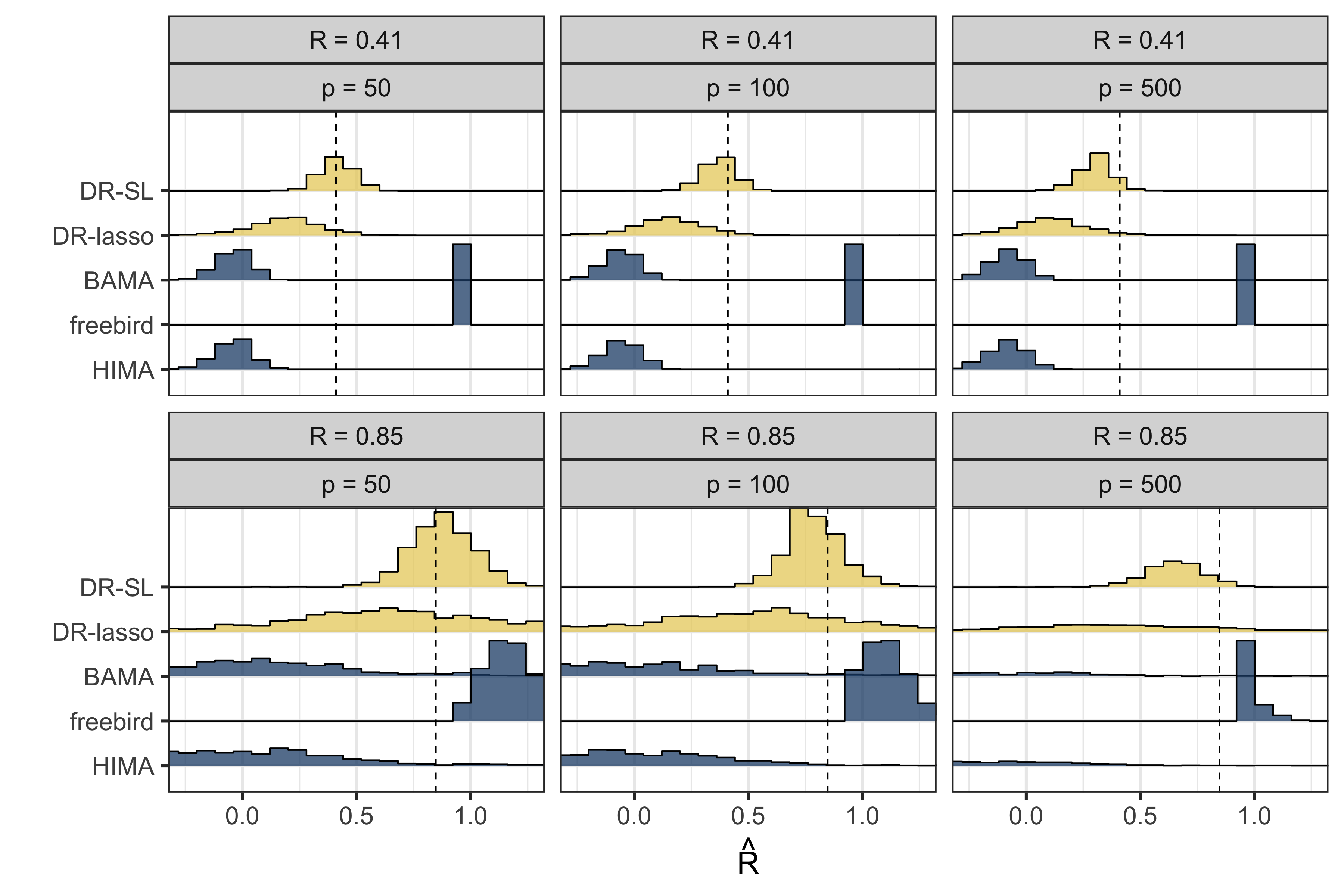}
\end{tabular}\vspace{0.2in}
\caption{Distribution of estimates of $\Rhat$ in first data-generating mechanism. Lighter shaded histograms represent distribution of the proposed estimators (``DR-SL" and ``DR-lasso"); darker shaded histograms represent the distribution of the comparison estimators (``BAMA", ``freebird", and ``HIMA"). The true value of $R$ is given as a vertical dotted line.}
\label{r-pl2}
\end{figure}

This data-generating mechanism was not as well approximated by linear models, in part because of the interactions between $X_1$ and $X_2$. Thus, the linear model-based estimates of HIMA, BAMA, and freebird all performed poorly (see Figure \ref{r-pl2}). HIMA and BAMA typically had $\Rhat$ values around 0 and lower, while freebird again estimated $R$ to be 1 or above (the 2.5\% quantile of the empricial distribution of freebird estimates was at 1 in all settings). On the other hand, the two proposed approaches gave more reasonable if imperfect estimates. When the dimensionality of the surrogates was much smaller than the sample size ($p = 50$), the DR-SL estimator gave nearly unbiased results, with median $\Rhat$ values of 0.88 (when $R = 0.85$) and 0.41 (when $R = 0.45$), as it was able to adapt to the nonlinearity in the data-generating mechanism. As the dimensionality of the surrogate grew, the Super Learner estimates of nuisance functions began to lean more heavily on linear models (like the lasso), and performance degraded: median $\Rhat$ values were 0.78 ($R = 0.85, p = 100$), 0.37 ($R = 0.41, p = 100$), 0.65 ($R = 0.85, p = 500$), 0.31 ($R = 0.41, p = 500$). The DR-lasso estimator resulted in performance between HIMA/BAMA and DR-SL, typically underestimating $R$ by more than DR-SL but less than HIMA/BAMA. CI coverage was strong for DR-SL when $R = 0.41$ (97\%, 96\%, 98\% for $p = 50, 100, 500$, respectively) but was much weaker when $R = 0.85$ (91\%, 74\%, 69\% for $p = 50, 100, 500$, respectively). This subpar coverage appeared to be due to the bias of the estimator and not due to the width of confidence intervals. When CIs were adjusted for the empirical bias of the DR-SL estimator (by subtracting the empirical bias from both sides of CIs), nominal coverage was observed (96\%, 97\%, 95\% coverage for $p = 50, 100, 500$, respectively).

\section{Discussion \label{discussion}}

We have proposed a very general approach to evaluating surrogate markers which can be applied in randomized experiments or in observational studies and can be used regardless of the dimensionality of either surrogates or any covariates needed to control for confounding. Our approach is nonparametric in that we have defined the proportion of the treatment effect explained by the surrogates without reference to any models, and we have shown how machine learning approaches like the super learner may be used to very flexibly estimate nuisance functions. Our simulation results suggest that our the Super Learner estimator (the estimator we called DR-SL) outperforms competing methods even when the underlying data-generating mechanism is linear and still gives reasonable results even when high dimensional linear models are mis-specified.

We note that the proposed estimator of the PTE, $\Rhat_\bS$, is ``doubly-robust" in the sense of \textit{rate double robustness} discussed in \cite{smucler2019unifying}, where the rates of two estimators (for us, the estimators of $\mu_a$ and $\pi$ as well as the estimators of $m_a$ and $e$) may compensate for one another -- i.e., a slightly slower rate of convergence for $\pi$ may be compensated for by a slightly faster rate of convergence for $\mu_a$ so long as their product is $o_p(\nnhalf)$. In high dimensions, the approach of \cite{smucler2019unifying}, based on $\ell_1$-penalized regression, could be used to additionally obtain the slightly different property of \textit{model double robustness} where one of the estimators may converge to a different value besides the true parameter. Our development here, which we chose to be quite general and applicable in both low- and high-dimensional settings, could easily be adapted to use the estimation technique of \cite{smucler2019unifying} in high dimensions to recover model double robustness.

Our approach here is intimately tied to previous approaches for evaluating surrogate markers. The approaches in \cite{agniel2020evaluation} and in \cite{parast2016robust} can be seen to be similar to a version of our approach where the average treatment effect among controls is estimated under further assumptions. In \cite{agniel2020evaluation}, they further require strict randomization of treatment,
and they assume that $\bS$ is a realization of a smooth continuous function. \cite{parast2016robust} make similar assumptions but take $S$ to be a scalar surrogate. They estimate a version of \eqref{om} among controls: $\Delta_{\bS} = \Ebb\left\{\mu_1(\bS) - \mu_0(\bS) | A = 0\right\}$ using kernel smoothing and taking advantage of the fact that treatment is randomized. Their estimates have the form
\[
\Deltahat_{\bS} = n_0\inv\sum_{A_i = 0}\left\{\muhat_1(\bS_i) - Y_i\right\}
\]
where $n_0 = \sumin I\{A_i = 0\}$ and $\muhat_1(\cdot)$ is estimated via kernel smoothing (possibly after dimension reduction). Our approach here could easily be adapted to estimate a similar quantity by using methods for doubly robust estimation of the average treatment effect on the treated \cite{shu2018improved, moodie2018doubly} by, for example,
\[
\Deltahat_{\bS} = n\inv\sumin\left\{A_i\frac{1-\pihat(\bS_i)}{\pihat(\bS_i)} - (1-A_i)\right\}\left\{\muhat_1(\bS_i) - Y_i\right\}
\]
where $\pihat(\bs) = \Phat(A_i | \bS_i = \bs)$ may be estimated using a model similar to the one used for $\muhat_1(\bs)$. Furthermore, our cross-fitting approach could be used to facilitate the use of machine learning methods in the estimation of $\muhat_1(\cdot), \pihat(\cdot)$, to include covariates to control for confounding, or to simplify or strengthen asymptotic results -- e.g., results for the kernel estimator used in \cite{agniel2020evaluation} obtain rates of convergence of $\nnhalf$ only under very limited technical conditions.

Our proposed robust approach to examine the strength of a high-dimensional surrogate will be useful in practice as there is increased attention on utilizing high-dimensional data that can be measured earlier or with less cost compared to primary outcomes, in an effort to make conclusions about treatments and interventions sooner. We include software to implement our proposed methods in the R package \texttt{crossurr} available at \texttt{github.com/denisagniel/crossurr}. 

\bibliographystyle{plain}
\bibliography{bibliography}

\section{Appendix}
\subsection{Details for example depicted in Figure 1}\label{fig1-sim}
In this small simulation, we let $n = n^* = 1000$, and we allowed $A$ to be randomized with $P(A = 1) = 0.5$. The rest of the data were generated as follows: $W = \delta A - 1 + \epsilon_w$, $S = \delta W + \epsilon_s, Y = 1.5A + W + \epsilon_y$. All three errors were independently generated as $\epsilon_w \sim N(0, 1), \epsilon_s \sim N(0, 1), \epsilon_y \sim N(0, 0.25\sqrt{\delta + 1}).$ The surrogate $\psi_a(s)$ was taken to be an estimate of $E(Y | A = a, S = s)$ obtained as the predicted value from a least squares fit of $Y$ on $S$ in treatment group $A = a$ from study 1. Figure 1 depicts $\delta$ values of 0 (PTE 0), 0.25 (PTE 0.2), 0.5 (PTE 0.33), and 3 (PTE 0.75).

\subsection{Regularity conditions}\label{assump-reg}
We follow the results of \cite{Chernozhukov2017} and require minimal regularity conditions on the distribution of $\bO$. Specifically, we 
require that for some $q_1, q_2 > 4$ $C_1, C_2 > 0$, $\Ctilde = \min{\{C_1, C_2\}}$, and $a = 0,1$
\[
\left[E\{\left|\mu_a(\bX, \bS)\right|^{q_1}\}\right]^{1/q_1} \leq C_1, \left[E\{\left|m_a(\bX)\right|^{q_2}\}\right]^{1/q_2} \leq C_2,
\]
and further that $\left\{E(|Y|^{q_1})\right\}^{1/q_1} \leq \Ctilde$. We finally assume that errors are well behaved, for $c_1, c_2 > 0, a = 0, 1$: 
\begin{align*}
    &E\left[\{Y - \mu_A(\bX, \bS)\}^2 | \bX, \bS\right] \leq C_1,
    E\left[\{Y - m_A(\bX)\}^2 | \bX\right] \leq C_2\\
    &E\left[\{Y - \mu_A(\bX, \bS)\}^2 \right] > c_1, 
    E\left[\{Y - m_A(\bX)\}^2 \right] > c_2\\
    &E\left[\{A - \pi(\bX, \bS)\}^2 \right] > c_1, 
    E\left[\{A - e(\bX)\}^2 \right] > c_2.
\end{align*}

\subsection{Justification of perturbation resampling}\label{ptb-app}
In this section, we will show that the limiting distribution (conditional on the observed data) of the resampled $\Rhat_{\bS b}^*$ is the same as the limiting distribution of $\Rhat_\bS$. Recall that
\begin{align*}
    \Deltahat_b^* = n\inv\sumin \Gsc_{ib}\phihat_{1i}\\
    \Deltahat_{\bS b}^* = n\inv\sumin \Gsc_{ib}\phihat_{2i},
\end{align*}
and let $\bthetahat_b^* = (\Deltahat_b^*, \Deltahat^*_{\bS b})\trans.$
Therefore, $E(\bthetahat_b^* | \Osc) = \bthetahat$ and $\cov(\bthetahat_b^* | \Osc) = n^{-2}\sumin\bphihat_i\bphihat_i\trans = n\inv\Sigmahat$.  Therefore, 
\begin{align*}
    \left.\sqrt{n}\Sigmahat^{-\frac{1}{2}}(\bthetahat^*_b - \bthetahat)\right|\Osc  \rightarrow N(\bzero, I),
\end{align*}
and the delta method yields the result 
\begin{align*}
    \left.\sqrt{n}\sigmahat\inv(\Rhat_{\bS b}^* - \Rhat_\bS)\right|\Osc \rightarrow N(0, 1).
\end{align*}

\subsection{Implementation details for simulations}
Here we provide further details on the implementation of the proposed estimator in simulation.

In using Super Learner to estimate nuisance functions, we specified the following candidate learners. For $\mu_1(\bX, \bS), \mu_0(\bX, \bS)$, we included the following learners: the mean, the lasso, ridge regression, ordinary least squares, support vector machines (SVM), and random forests. For $\pi(\bX, \bS)$, we included: the mean, the lasso, logistic regression, linear discriminant analysis, quadratic discriminant analysis, SVM, and random forests. For $m_1(\bX), m_0(\bX)$, we additionally considered a generalized linear model including all interaction terms, step-wise regression including all interaction terms, and regression trees. And for $e(\bX)$, we additionally included logistic regression with all interaction terms, classification trees, and k-nearest neighbors. We took $K = 4$ for cross-fitting, and we used the default cross-validation procedure to select the tuning parameters for the candidate learners and the Super Learner. 

For the DR-lasso estimator, we used the default cross-validation to select the tuning parameters. 

\end{document}

%% file: GrandMacros.tex
\def\bzero{{\bf 0}}
\def\bone{{\bf 1}}
\def\be{{\bf e}}

\def\bs{{\bf s}}

\def\bx{{\bf x}}

\def\bO{{\bf O}}

\def\bS{{\bf S}}

\def\bX{{\bf X}}

\def\thick#1{\hbox{\rlap{$#1$}\kern0.25pt\rlap{$#1$}\kern0.25pt$#1$}}
\def\balpha{\boldsymbol{\alpha}}
\def\bbeta{\boldsymbol{\beta}}
\def\bgamma{\boldsymbol{\gamma}}

\def\btheta{\boldsymbol{\theta}}

\def\bphi{\boldsymbol{\phi}}

\def\smbalpha{\boldsymbol{{\scriptstyle{\alpha}}}}

\def\ehat{{\widehat e}}

\def\mhat{{\widehat m}}

\def\Ehat{{\widehat E}}

\def\Phat{{\widehat P}}

\def\Rhat{{\widehat R}}

\def\Ctilde{{\widetilde C}}

\def\muhat{{\widehat\mu}}

\def\pihat{{\widehat\pi}}

\def\sigmahat{{\widehat\sigma}}

\def\phihat{{\widehat\phi}}

\def\Deltahat{{\widehat\Delta}}

\def\Sigmahat{{\widehat\Sigma}}

\def\bthetahat{{\widehat\btheta}}

\def\bphihat{{\widehat\bphi}}

\def\smbalpha{\widehat{\smbalpha}}

\def\hbar{\bar{ h}}

\def\Ybar{\bar{ Y}}

\def\Csc{{\cal C}}

\def\Fsc{{\cal F}}
\def\Gsc{{\cal G}}

\def\Isc{{\cal I}}

\def\Msc{{\cal M}}

\def\Osc{{\cal O}}

\def\etal{{\em et al.}}
\def\transpose{{\sf \scriptscriptstyle{T}}}

\def\half{\frac{1}{2}}

\def\nnhalf{n^{-\half}}

\def\sumin{\sum_{i=1}^n}

\def\trans{^{\transpose}}
\def\inv{^{-1}}

\def\cov{\mbox{cov}}

\def\mybox#1{\vskip1mm \begin{center}
        \hspace{.0\textwidth}\vbox{\hrule\hbox{\vrule\kern6pt
\parbox{.9\textwidth}{\kern6pt#1\vskip6pt}\kern6pt\vrule}\hrule}
        \end{center} \vskip-5mm}
\def\lboxit#1{\vbox{\hrule\hbox{\vrule\kern6pt
      \vbox{\kern6pt#1\vskip6pt}\kern6pt\vrule}\hrule}}

\def\thickboxit#1{\vbox{{\hrule height 1mm}\hbox{{\vrule width 1mm}\kern6pt
          \vbox{\kern6pt#1\kern6pt}\kern6pt{\vrule width 1mm}}
               {\hrule height 1mm}}}

\def\fat#1{\hbox{\rlap{$#1$}\kern0.25pt\rlap{$#1$}\kern0.25pt$#1$}}